\documentclass[twocolumn,showpacs,superscriptaddress,preprintnumbers,prl]{revtex4-1}
\usepackage{graphicx,amsmath,amssymb,dcolumn,bm}
\usepackage{fixmath}
\usepackage{epic,eepic}
\usepackage{slashed}

\allowdisplaybreaks[2]
\usepackage{calrsfs}
\DeclareMathAlphabet{\pazocal}{OMS}{zplm}{m}{n}
\def\ds{\displaystyle}
\begin{document}
\title{First determination of fragmentation functions\\
in an exotic-hadron candidate}
\author{S. Kumano}

\affiliation
{Quark Matter Research Center,
    Institute of Modern Physics, Chinese Academy of Sciences,\\
    Lanzhou, 730000, China,\\
 Southern Center for Nuclear Science Theory,
    Institute of Modern Physics, Chinese Academy of Sciences,\\
    Huizhou, 516000, China}
\affiliation
{KEK Theory Center, Institute of Particle and Nuclear Studies, KEK,
Oho 1-1, Tsukuba, 305-0801, Japan}
% \date{\today}
% \date{November 25, 2025}
\date{September 7, 2026}
%%%%%%%%%%%%%%%%%%%%%%%%%%%%%%%%%%%%%%%%%%%%%%%%%%%%%%%%%%%%%%%%%%%%%%%%%%%%%%%%

%%%%%%%%%%%%%%%%%%%%%%%%%%%%%%%%%%%%%%%%%%%%%%%%%%%%%%%%%%%%%%%%%%%%%%%%%%%%%%%%
\begin{abstract}
In recent years, there are experimental reports on exotic-hadron candidates,
which have different quark configurations from ordinary $q\bar q$
and $qqq$ constituents. However, it is not easy to confirm 
their exotic nature from global observables such as masses, spins, 
parities, and decay widths. At high energies, internal quark and gluon
configurations could become more apparent because hadrons should be described
by fundamental degrees of freedom of quarks and gluons
in quantum chromodynamics.
One of such possibilities is to use the fragmentation functions (FFs). 
In this work, accurate FFs of an exotic hadron candidate $f_0$(980) are 
determined for the first time by an global analysis of experimental data
on $e^+ + e^- \to f_0 (980)+X$ with recent precise measurements
of the Belle collaboration. From the global analysis, we found 
that their second moments have a relation
$M_u = M_d \ll M_s \sim M_g $ for up-quark, down-quark, 
strange-quark, and gluon FFs. Furthermore, the function $D_s^{f_0}(z)$ is 
distributed in the larger-$z$ region in comparison with
the functions $D_{u}^{f_0}(z)$, $D_{d}^{f_0}(z)$, and $D_g^{f_0}(z)$. 
These facts support that the $f_0 (980)$ has the $s\bar s$ configuration
at high energies.
This is a new finding that $f_0 (980)$ should be considered
mainly as the $s\bar s$ state, which is different 
from our usual understanding as a tetraquark (or $K\bar K$) hadron
from low-energy studies.
Our results could indicate the transition of the internal configuration 
picture that $f_0 (980)$ looks like a $q\bar q$ state
at high energies although it is described by tetra-quark or $K\bar K$
molecule state at low energies.
It sheds light on a new direction in exotic hadron physics.
\end{abstract}
\maketitle

%%%%%%%%%%%%%%%%%%%%%%%%%%%%%%%%%%%%%%%%%%%%%%%%%%%%%%%%%%%%%%%%%%%%%%%%%%%%%%%%
\noindent
{\it Introduction}: 
Many hadrons are described by quark models by assuming
ordinary $qqq$ and $q\bar q$ configurations as proposed
in the original quark model of 1964.
In recent years, there have been reports on exotic hadron candidates
($e.g.$ $qq\bar q \bar q$ and $qqqq\bar q$),
which are different from these basic configurations. 
They are called exotic hadrons \cite{pdg-2024}.
Recent investigations focus on heavy-quark systems.
On the other hand, there are longstanding exotic hadron candidates
in the 1 GeV mass region, and $f_0$(980) is such an example
\cite{pdg-2024,cik-1993,sk-2015}. It is considered as a tetrqquark
or $K\bar K$-molecule-like hadron from low-energy studies
\cite{pdg-2024,cik-1993,sk-2015,sk-2015}.

In general, it is difficult to identify internal configurations
of exotic hadron candidates undoubtedly by global observables 
such as masses, spins, parities, and decay widths.
On the contrary, high-energy reaction observables could be
appropriate for such identifications because 
quarks and gluons are explicit degrees of freedom
at high energies. Along this line, we proposed 
fragmentation functions (FFs) \cite{hkos-2008}, 
constituent counting rule \cite{const-conting-2013,const-conting-2016}, 
and generalized parton distributions \cite{gpd} as 
such high-energy observables.

A fragmentation function $D_i^h$ indicates
the probability for producing the hadron $h$ from the parton $i$,
which is up-, down-, strange-, charm-, bottom-quark, or gluon.
Now, the FFs are relatively well determined by global analyses
for major hadrons such as proton, pion, kaon, and other light hadrons
\cite{FFs-global-1,FFs-global-2}.
In general, the FFs are classified into 
favored and disfavored functions as the parton distributions
are classified into valence- and sea-quark distributions.
The favored fragmentation indicates a fragmentation 
from a quark or an antiquark which exists within a hadron 
as a constituent in a quark model, and the disfavored means 
a fragmentation from a sea quark. Therefore, 
the favored and disfavored functions reflect internal
quark and gluon configuration, which can be used
for finding whether or not the hadron is really exotic.
If the FFs are determined for exotic hadron candidates
by a global analysis of experimental data, we should be 
able to determine their internal structure.
In particular, if the $f_0$(980) has a tetrqquark
or $K\bar K$-molecule structure, the favored FFs
should be very different from the FFs of 
an ordinary $q\bar q$ meson.

In this work, we report a global analysis of $f_0$(980) fragmentation
function data for determining its internal configuration.
At the stage of 2008 \cite{hkos-2008}, this kind of analysis 
was too early because there were not so many experimental data 
and their errors are very large on the measured fragmentation functions.
The situation changed recently because accurate measurements
have been done for light hadrons including $f_0$(980)
by the Belle Collaboration in 2025 \cite{KEKB-2025}.
The high-intensity KEK-B factory is suitable 
for such precision studies.
With these recent KEK-B data, it became realistic 
to propose accurate FFs of an exotic hadron candidate $f_0$(980)
for the first time, and it enables us to find 
its internal exotic structure from the FFs.

The purpose of this work is to report analysis results
including the KEK-B data for determining the internal configuration 
of $f_0$(980). By the determination of the $f_0 (980)$ FFs,
we expect that it is possible to distinguish
among possible internal configurations,
a nonstrange $q\bar q$ ($(u\bar u + d\bar d)/\sqrt{2}$),
a strange $q\bar q$ ($s\bar s$),
a tetraquark or $K\bar K$ ($(u\bar u s\bar s +d\bar d s\bar s)/\sqrt{2}$),
and a glueball ($gg$).
In the following, we explain the analysis method
of the $f_0$(980) FFs and its results to discuss
the internal configuration picture.
As shown in the following, we found that $f_0$(980) looks like
an $s\bar s$ state, which is contradictory to the usual understanding
that it is a tetraquark or $K\bar K$ meson.
In the end of this paper, we suggest a possible explanation
how these seemingly inconsistent pictures could be reconciled
without a contradiction.

%%%%%%%%%%%%%%%%%%%%%%%%%%%%%%%%%%%%%%%%%%%%%%%%%%%%%%%%%%%%%%%%%%%%%%%%%%%%%%%%
\vspace{0.00cm}
\noindent
{\it Analysis method}: 
The fragmentation functions are measured by high-energy 
hadron production processes. A typical one 
is the inclusive hadron production in the electron-positron collision
($e^+e^- \rightarrow hX$). The fragmentation function $F^h(z,Q^2)$ is
defined by the $e^+e^-$ total cross section $\sigma_{tot}$
and the $e^+e^- \rightarrow hX$ cross section 
$\sigma (e^+e^- \rightarrow hX)$ as
$F^h(z,Q^2) = d\sigma (e^+e^- \rightarrow hX) / dz / \sigma_{\rm tot}$
\cite{esw-book}
where the variable $z$ is defined as $z = E_h/(\sqrt{s}/2)$
with the hadron energy $E_h$ and the center-of-mass energy 
$\sqrt{s} = \sqrt{Q^2}$ of $e^+ e^-$.
%%%%%
The fragmentation occurs from primary partons,
so that it is expressed theoretically
by the sum of their contributions
$F^h(z,Q^2) = \sum_i C_i(z,\alpha_s) \otimes D_i^h (z,Q^2) $,
where $\otimes$ indicates the convolution integral
$f (z) \otimes g (z) = {\int}^{1}_{z} dy f(y) g(z/y)/y$.
Here, $D_i^h(z,Q^2)$ is the fragmentation function
of the hadron $h$ from a parton $i$ ($=u,\ d,\ s,\ \cdot\cdot\cdot,\ g$),
$C_i(z,\alpha_s)$ is a coefficient function to include
perturbative QCD corrections, 
and $\alpha_s$ is the running coupling constant.

From theoretical and experimental studies at low energies, it is likely 
that $f_0$(980) is a tetraquark or $K\bar K$ molecule \cite{pdg-2024}.
However, let us consider all the possible configurations
without prejudice. Then, a scalar meson in the 1 GeV region could be 
formed as a nonstrange $q\bar q$ meson ($(u\bar u+d\bar d)/\sqrt{2}$),
a strange $q\bar q$  ($s\bar s$),
tetraquark or $K\bar K$ ($(u\bar u s\bar s+d\bar d s\bar s)/\sqrt{2}$),
or a glueball ($gg$) \cite{cik-1993}.
%%%%%
In all these configurations, the quark $u$, $d$, $\bar u$, and $\bar d$
are contained in the same fraction, which indicates
that the parameters of these FFs could be assumed to be the same.
The quark $s$ and $\bar s$ are also same, so that the common FFs
could be used for both $s$ and $\bar s$.
Therefore, we express the fragmentation functions
$ D_{u}^{f_0} (z,Q_0^2) = D_{\bar u}^{f_0} (z,Q_0^2)
 = D_{d}^{f_0} (z,Q_0^2) =  D_{\bar d}^{f_0} (z,Q_0^2)$,
$D_{s}^{f_0} (z,Q_0^2)$, $D_{g}^{f_0} (z,Q_0^2)$,
$D_{c}^{f_0} (z,m_c^2)$, and $D_{b}^{f_0} (z,m_b^2)$
in terms of parameters, 
which are determined by a $\chi^2$ analysis of the data for
$e^++e^- \rightarrow f_0+X$. 
Each function is expressed by the three parameters
$N_{i}$, $\alpha_{i}$, and $\beta_{i}$ as
$ D_i (z,Q_0^2) = N_{i} z^{\alpha_i} (1-z)^{\beta_i}$
where $i=u$ (or $d$), $s$, $c$, $b$, and $g$.
%%%%%
This parametrization can account both the favored and
disfavored FFs and the disfavored ones are suppressed
at large $z$ according to Figs.\,1 and 2 of Ref.\,\cite{hkos-2008}.
%%%%%
The initial scale is $Q_0^2=1$ GeV$^2$, and $m_c$ 
and $m_b$ are taken $m_c$=1.43 GeV and $m_b$=4.3 GeV.
$Q^2$ variations to the experimental $Q^2$ points are calculated 
by the Dokshitzer-Gribov-Lipatov-Altarelli-Parisi equation
\cite{FFs-evol-2012}.
The second moment of a fragmentation function $D_i (z,Q^2)$
is given by the integral 
$M_i = \int_0^1 dz z D_i (z,Q^2)$.
The parameters $N_i$ and $M_i$ are related by
$N_i^h = M_i / B(\alpha_i^h+2, \beta_i+1)$
with the beta function $B(\alpha_i+2, \beta_i+1)$.
The $\chi^2$ analysis is done in the next-to-leading order (NLO)
of $\alpha_s$ and uncertainties of the FFs are calculated
by the Hessian method.

%%%%%%%%%%%%%%%%%%%%%%%%%%%%%% table %%%%%%%%%%%%%%%%%%%%%%%%%%%%%
\begin{table}[t]
\vspace{-0.18cm}
\caption{Experimental collaborations, laboratories, references, 
     center-of-mass energies in GeV, numbers of data points,
     kinematical variables, cross sections are listed 
     for used data sets of 
     $e^+ +e^- \rightarrow f_0 (980) +X$.}
\label{tab:exp-f0}
\begin{ruledtabular}
\centering
\begin{tabular}
{l@{\extracolsep{10ptplus1fil}}c@{\extracolsep{10ptplus1fil}}c
@{\extracolsep{10ptplus1fil}}c@{\extracolsep{10ptplus1fil}}c
@{\extracolsep{10ptplus1fil}}c@{\extracolsep{10ptplus1fil}}c}
Exp.  &  Lab. \ & Ref.  & \hspace{-0.20cm} $\sqrt{s}$\ \, & \hspace{-0.30cm} Data \# \hspace{-0.20cm}
            & Var. & \hspace{-0.20cm} Cross \\
\hline
HRS         &  SLAC   & \cite{hrs86}    & \hspace{-0.20cm} 29 \ \ \ \  & \, \hspace{-0.20cm} 4 
            & $z_p$      
            & \ \ \ \hspace{-0.20cm} $\ds\frac{s}{\beta}\frac{d\sigma}{dz}$  \\
DELPHI \hspace{-0.25cm}    
            &  CERN   & \cite{delphi9599} & \hspace{-0.20cm} 91.2  \,  &  \hspace{-0.20cm} 11  
            & $z_p$      
            & \hspace{-0.20cm} $\ds\frac{1}{\sigma_{\rm tot}}\frac{d\sigma}{dz_p}$ \\
OPAL        &  CERN   & \cite{opal98}   & \hspace{-0.20cm} 91.2  \,    & \, \hspace{-0.20cm} 8 
            & $\bar z_p$ 
            & \hspace{-0.20cm} $\ds\frac{1}{\sigma_{\rm tot}}\frac{d\sigma}{d\bar z_p}$  \\
Belle       &  KEK \  & \cite{KEKB-2025}  & \, \hspace{-0.20cm} 10.58 \hspace{0.10cm} & \hspace{-0.20cm} 40  
            & $z_p$ 
            & \ \ \ \ \ \ \hspace{-0.20cm} $\ds\frac{d\sigma}{dz_p}$ \\
\hline
total       &         &                 &             &  63 
            &  
            &  \\
\end{tabular}
\end{ruledtabular}
\vspace{-0.40cm}
\end{table}
%%%%%%%%%%%%%%%%%%%%%%%%%%%%% table %%%%%%%%%%%%%%%%%%%%%%%%%%%%%%

%%%%%%%%%%%%%%%%%%%%%%%%%%%% figure %%%%%%%%%%%%%%%%%%%%%%%%%%%%
\begin{figure}[b]
 \vspace{-0.60cm}
\begin{center}
   \includegraphics[width=7.0cm]{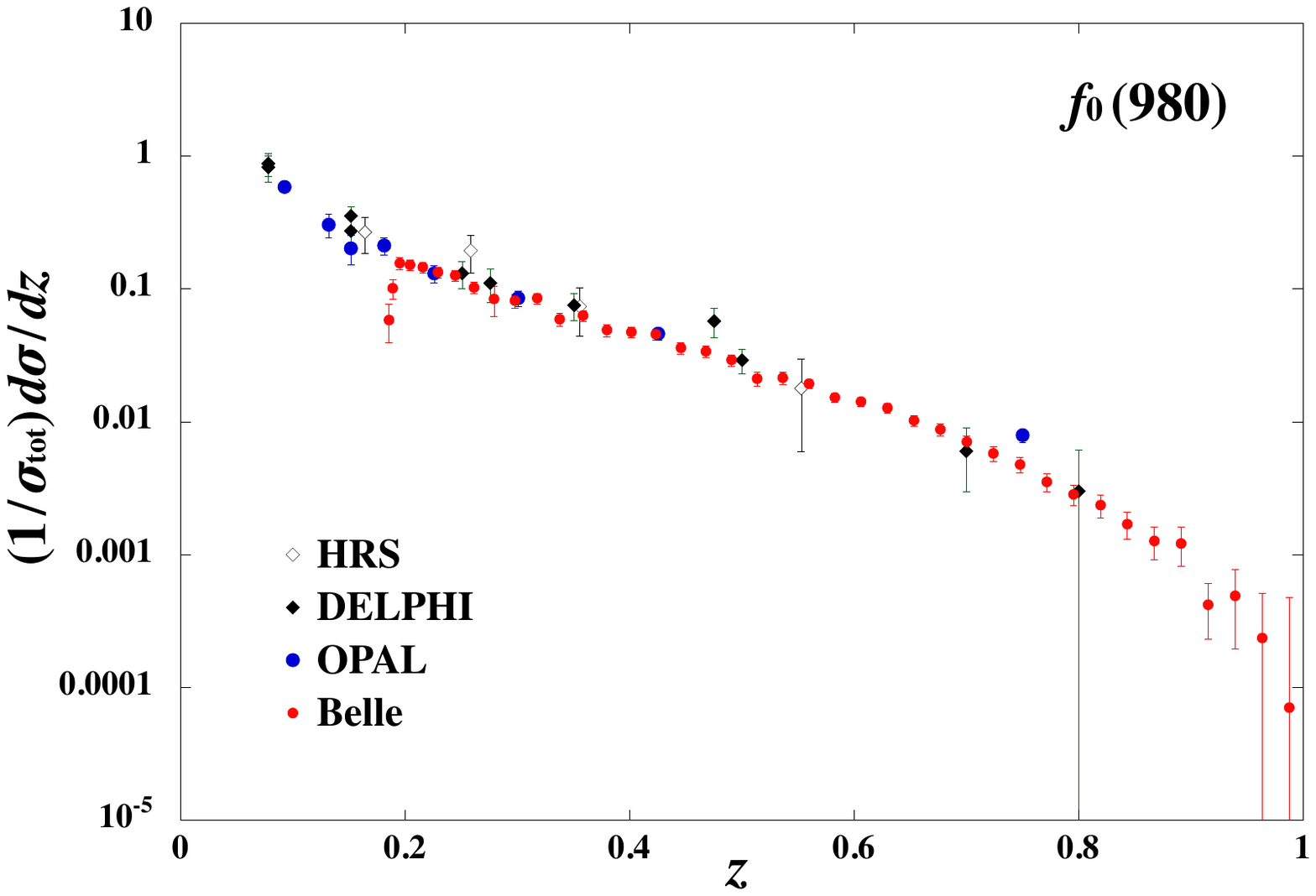}
\end{center}
\vspace{-0.9cm}
\caption{Fragmentation-function data are shown 
for $f_0$(980) in the form of 
$F^{f_0} = d\sigma (e^+ e^- \to f_0 X) / dz /\sigma_{\rm{tot}}$.}
\label{fig:f0_980-data}
\vspace{-0.00cm}
\end{figure}
%%%%%%%%%%%%%%%%%%%%%%%%%%%% figure %%%%%%%%%%%%%%%%%%%%%%%%%%%%

Past $e^++e^- \rightarrow f_0+X$ experiments 
\cite{KEKB-2025,hrs86,delphi9599,opal98}
are listed in 
Table \ref{tab:exp-f0}. Different kinematical variables
are used in showing the data depending on 
the experimental collaborations, and they are defined by
$ z = z_E = 2 E_h /\sqrt{s}$, 
$ z_p = 2 p_h / \sqrt{s-4 m_h^2}$, 
$ \bar z_p = 2 p_h / \sqrt{s}$,
where $p_h$ and $m_h$ are the $f_0$ momentum and its mass, respectively.
The experimental cross sections are shown by slightly different 
quantities as shown in Table \ref{tab:exp-f0}, 
so that they are converted to the same FFs 
$F^{f_0}(z,Q^2)$ for calculating $\chi^2$
in comparison with theoretical FFs.
The results are shown in Fig.\,\ref{fig:f0_980-data}.
Although $Q^2$ values are different between the experimental groups,
four experimental data sets are consistent with each other.
It is obvious from this figure that the situation of
the $f_0 (980)$ FFs was improved significantly recently because
of the Belle data in 2025. The previous HRS, DELPHI, and OPAL data
are taken in scattered kinematical points with large errors,
so that one can imagine that the determination of 
the FFs was difficult only from these data.
Including the precise Belle data in the wide kinematical region of $z$,
we expect that the $f_0 (980)$ FFs can be determined accurately.

%%%%%%%%%%%%%%%%%%%%%%%%%%%%%%%%%%%%%%%%%%%%%%%%%%%%%%%%%%%%%%%%%%%%%%%%%%%%%%%%
\vspace{0.00cm}
\noindent
{\it Results}: 
By the $\chi^2$ analysis of the data in Fig.\,\ref{fig:f0_980-data},
the optimum FFs are determined with 
$\chi^2/\rm{d.o.f.}=1.43$,
where we fixed two parameters $\beta_c$ and $\beta_b$, which control 
the function forms of $D_c^{f_0} (z)$ and $D_b^{f_0} (z)$ at large $z$,
in the charm and bottom functions as $\beta_c = \beta_b =10$.
However, we do not fix the parameters of the up (down), strange,
and gluon functions, because they are crucial 
for determining the internal configuration of $f_0 (980)$.
Therefore, the total number of the parameters is 13.
As seen in Fig.\,\ref{fig:f0_980-data}, it is almost impossible
to determine the FFs of $f_0 (980)$ without the Belle data. 
In fact, we obtained huge uncertainty bands for the $f_0 (980)$
FFs in 2008 \cite{hkos-2008}. However, the Belle data are accurate enough
and they enable us to determine the FFs precisely.

%%%%%%%%%%%%%%%%%%%%%%%%%%%% figure %%%%%%%%%%%%%%%%%%%%%%%%%%%%
\begin{figure}[b]
 \vspace{-0.40cm}
\begin{center}
   \includegraphics[width=7.0cm]{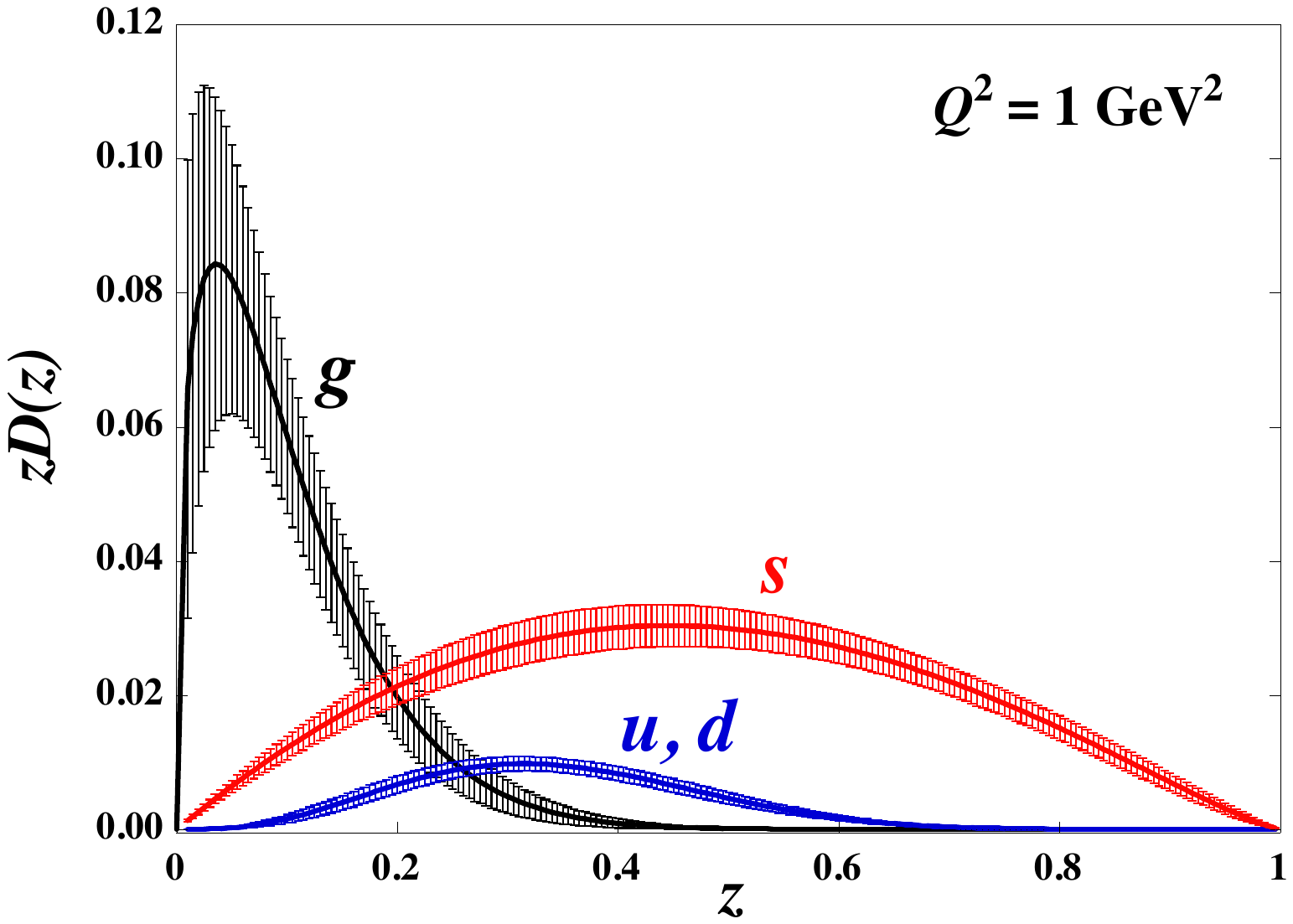}
\end{center}
\vspace{-0.9cm}
\caption{Determined fragmentation functions for $f_0$(980).}
\label{fig:f0_980-ffs}
\vspace{-0.00cm}
\end{figure}
%%%%%%%%%%%%%%%%%%%%%%%%%%%% figure %%%%%%%%%%%%%%%%%%%%%%%%%%%%

By the NLO $\chi^2$ analysis, the second moments 
of $D_u^{f_0}$ ($=D_d^{f_0}$), $D_s^{f_0}$, 
and $D_g^{f_0}$ are determined as
$ M_u = M_d  
      = 0.00329  \pm 0.00012 $, 
$ M_s = 0.01936  \pm 0.00041 $,
$ M_g = 0.01231  \pm 0.00074 $ 
at $Q^2=1$ GeV$^2$.
These values indicate a relation among these second moments as
\begin{align}
M_u = M_d \ll M_s \sim  M_g .
\label{eqn:moment-relation}
\end{align}
This relation indicates that the $u$ and $d$ quarks are not
the main components of $f_0 (980)$, although 
they should be the main ingredients in the tetraquark
or $K\bar K$ picture from the low energy studies.
The other parameters are obtained as
$\alpha_g = -0.559 \pm 0.071$,
$\beta_g  = 11.86  \pm 0.94$,
$\alpha_u = 2.33 \pm 0.11$,
$\beta_u  = 7.266 \pm 0.060$,
$\alpha_s = 0.2623 \pm 0.0023$,
$\beta_s  = 1.276 \pm 0.043$,
$M_c      = 0.00073 \pm 0.00012$,
$\alpha_c = 3.557 \pm 0.073$,
$M_b      = 0.000983 \pm 0.000074$, and
$\alpha_b = 14.53 \pm 0.074$.
%%%%%
In the pion case, according to a recent global analysis
(J. Gao {\it et al}., 2025 \cite{FFs-global-2}),
the relation is given as
$M_{u,\bar d}^{\pi^+} = 0.481 \gg  M_g^{\pi^+} = 0.102,
\ M_{s,\bar s}^{\pi^+} = 0.234, M_{d, \bar u}^{\pi^+} = 0.119$
in the NLO analysis at $Q^2=(1.4)^2$ GeV$^2$, 
which indicates that the $u$ and $\bar d$ 
are the main components in $\pi^+$ as suggested 
by the quark model. 

The determined FFs $D_u^{f_0} (z) =D_d^{f_0}(Z)$, $D_s^{f_0}(z)$, 
and $D_g^{f_0} (z)$ are shown at $Q^2=1$ GeV$^2$ 
in Fig.\,\ref{fig:f0_980-ffs}
with uncertain bands with $\Delta \chi^2 = 14.8$ for
the 13 parameter $\chi^2$ fit \cite{minuit}.
It is obvious from this figure that 
the FFs of $f_0 (980)$ are now determined accurately,
whereas the FFs of the 2008 analysis 
had huge uncertainty bands.
The $s$-quark function $D_s(z)$ is much larger
than the light-quark ones $D_u(z)$ and $D_d(z)$, and
it should be noted that $D_s(z)$ is distributed in 
the larger-$z$ region than $D_u(z)$ and $D_d(z)$.
We also notice that the gluon function $D_g(z)$ is distributed 
in the small-$z$ region, although the function
itself is larger than the $s$-quark one. 

Gluon and quark-flavor separations in the analysis of
the inclusive $e^+ e^-$ data are explained in Ref.\,\cite{FFs-global-1} (2016).
Because the DELPHI/OPAL energy 91.2 GeV and the Belle energy 10.58 GeV
are well separated, the gluon FF is constrained mainly 
by the scaling violation of the FF data.
Because of the differences between the electric charges 
and weak couplings, it is possible to separate 
the two functions $D_{u^+}$ and $D_{d^+} + D_{s^+}$,
aside from the heavy-quark FFs.
This separation means that the separation between 
$D_{u}=D_{d}$ and $D_{s}$ in this work,
because of $D_{u^+} = D_{d^+}$ in $f_0 (980)$.
To test effects of heavy-quark FFs, the analysis 
was done by changing the fixed parameters as
$\beta_c =\beta_b =5$; however, the moment relation
and the distributions are not significantly changed
for judging the internal structure.
%%%%%
It should be noted that accurate data are needed
in future at high energies especially with charm- 
and bottom-quark tagging to obtain a solid fit result on the FFs.

%%%%%%%%%%%%%%%%%%%%%%%%%%%%%% table %%%%%%%%%%%%%%%%%%%%%%%%%%%%%
\begin{table}[t]
\vspace{-0.22cm}
\caption{Possible $f_0(980)$ configurations 
and typical features of its fragmentation functions.}
\label{tab:f0-config}
\begin{ruledtabular}
\begin{tabular}{ccc}
Configuration   
                        & 2nd moment
                        & Function      \\
\hline
$(u\bar u+d\bar d)/\sqrt{2}$ 
                        & \,\, $M_s<M_u<M_g$ \ \ 
                        & $z_u^{\rm max}>z_s^{\rm max}$   \\ 
$s\bar s$ 
                        & $M_u  <   M_s \lesssim M_g$    
                        & $z_u^{\rm max}<z_s^{\rm max}$   \\  
$(u\bar u s\bar s+d\bar d s\bar s)/\sqrt{2}$ 
           & $M_u \sim M_s \lesssim M_g$
           & $z_u^{\rm max} \sim z_s^{\rm max}$   \\  
$gg$   
                        & $M_u \sim M_s < M_g$
                        & $z_u^{\rm max} \sim z_s^{\rm max}$   \\  
\end{tabular}
\end{ruledtabular}
\vspace{-0.50cm}
\end{table}
%%%%%%%%%%%%%%%%%%%%%%%%%%%%% table %%%%%%%%%%%%%%%%%%%%%%%%%%%%%%

General criteria are shown in Table \ref{tab:f0-config}
for judging internal configuration of $f_0 (980)$
from the FFs by using the second moments and
their functional forms \cite{hkos-2008}.
For example, if $f_0 (980)$ is an $s\bar s$ type,
a gluon is emitted from a strange quark (or antiquark) 
and the gluon splits into an $s\bar s$ pair
to form the $s\bar s$ bound state together
with the initial  strange quark (or strange antiquark).
This is a favored fragmentation process.
On the other hand, if an unfavored fragmentation
occurs from a $u$ quark, the $s \bar s$ from the gluon splitting
could form the $s\bar s$ bound state by radiating
another gluon. 
If it is the unfavored fragmentation from an initial gluon,
the  $s\bar s$ bound state is formed by the gluon splitting
and another gluon radiation to maintain the color singlet nature.
%%%%%
Because of the differences between these formation processes, 
there exist specific features in the second moments 
and the $z$-dependent functional forms.
In the favored fragmentation without the gluon emission,
the second moment of $D_s(z)$ is larger.
In addition, more energy is transferred to $f_0 (980)$
from the initial $s$ (or $\bar s$) quark, 
the function $D_s(z)$ is distributed at large $z$.
These relations are listed in the $s\bar s$ configuration part
of Table\,\ref{fig:f0_980-ffs}.
The other relations of this table are derived
by similar studies on the fragmentation processes to $f_0 (980)$.

One should note, however, that these criteria are very rough ideas,
especially on the functional form because quark-mass effects 
are not considered.
In fact, the determined $u$ and $\bar s$ FFs
are very different in $K^+$ \cite{FFs-global-1,FFs-global-2}
because the strange-quark mass is heavier,
although both are favored functions.
The first moments are different in favored 
and disfavored functions in global analyses,
for example, for pion and kaon; however, the difference 
between the functional shapes is not very clear at this stage. 
Therefore, these conditions should be 
considered as as a starting point 
to initiate much detailed theoretical studies 
including the mass effect.

Considering the criteria of Table \ref{tab:f0-config},
we decide the appropriate internal configuration of $f_0 (980)$.
The analysis results on the two factors, 
\vspace{-0.10cm}
\begin{itemize}
\setlength{\leftskip}{0.37cm}
\item the relation $M_u = M_d \ll M_s \sim M_g$
    in Eq.\,(\ref{eqn:moment-relation}), 
\vspace{-0.15cm}
\item the $z$-dependent functional forms 
of $D_u(z)$, $D_d(z)$, $D_s(z)$ and $D_s(z)$ 
in Fig.\,\ref{fig:f0_980-ffs}, 
\end{itemize}
\vspace{-0.10cm}
indicate that the $f_0$(980) looks like
an $s\bar s$ meson from its fragmentation functions 
from Table \ref{tab:f0-config}
because the $s$-quark moment $M_s$ is much larger than
the $u$- and $d$-quark ones $M_{u,d}$
and because the $D_s(z)$ is distributed in 
the larger-$z$ region than the other FFs.
%%%%%
In Eq.\,(\ref{eqn:moment-relation}), we find that
the gluon moment $M_{g}$ is as large as $M_s$; however,
it is not the major constituent by considering
the functional form distributed in the small-$z$ region. 
%%%%%
These facts indicate that the main constituent 
of $f_0 (980)$ is $s \bar s$, 
and the light quarks ($u$, $d$) and the gluon are not
the main components.
%%%%%%
In order to check the effect of the $Q_0^2 =1$ GeV$^2$ choice, 
the analysis was repeated by taking a larger value 
$Q_0^2 =4$ GeV$^2$. The obtained results are similar
($\chi^2/{\rm d.o.f.}=1.33, \ M_u = M_d = 0.00286 \pm 0.00034 \ll M_s = 0.01500 \pm 0.00041 
   \sim M_g = 0.0183 \pm 0.0015$) and the FFs are softer 
than the ones at $Q^2=1$ GeV$^2$ due to 
the $Q^2$ evolution \cite{FFs-evol-2012}, 
which supports the suggested picture for $f_0 (980)$.

This is the first clear indication on the internal structure
of the $f_0$(980) from an analysis of high-energy experimental data
on the fragmentation functions.
However, it makes us wonder about the difference from 
our ``common sense" from the low-energy studies 
that the $f_0$(980) is a tetraquark 
or $K\bar K$ molecule.
Properties of $f_0$(980) have been studied experimentally
at low energies for a long time, and it is 
our understanding that the $f_0$(980) is now considered as
a tetra-quark or $K\bar K$ molecule.
This picture is now confirmed, for example, 
by two-photon decay and radiative-decay widths of $f_0$(980)
\cite{pdg-2024}.
This picture and our analysis result of the FFs look like a paradox.

These seemingly contradictory results could be understood 
in the following way. The $\Lambda$(1405) is considered 
as a pentaquark or $\bar K N$ molecule from theoretical 
and experimental studies at low energies. 
The perturbative QCD indicates a specific feature, 
called the constituent counting rule, 
on cross sections of high-energy exclusive hadron processes.
This constituent counting analysis of $\Lambda$(1405)
photo-production data suggested the transition that 
$\Lambda$(1405) could be a pentqquark state at low energies
but it becomes an ordinary $qqq$ state at high energies
\cite{const-conting-2013}.
In the same way, $f_0$(980) could be a tetraquark state
at low energies and it becomes a $q\bar q$ state
at high energies. Our FF analysis supports this scenario:
\vspace{-0.20cm}
\begin{alignat}{2}
f_0 (980) = & \ \text{tetraquark state} & \ \to \ & q\bar q \ \text{state}
\nonumber\\[-0.10cm]
            & \text{(at low energy}      & \ \to \ & \text{high energy)} .
\nonumber
\label{eqn:f0-low-high-energies}
\vspace{-0.00cm}
\end{alignat}
\\[-0.60cm]
If this interpretation is right, this work could open 
a new view of hadrons and a new field of hadron physics. 
Exotic hadron candidates could be interpreted as 
``exotic" hadrons at low energies, whereas they 
could be viewed as ordinary hadrons 
with the $q\bar q$ and $qqq$ configurations at high energies.
Obviously, additional experimental confirmations are needed
to confirm this idea on the transition scenario.
The original naive quark model could be valid at high energies,
where quarks and gluons are explicit fundamental degrees 
of freedom of QCD, whereas some hadrons are considered as
exotic ones at low energies.

%%%%%%%%%%%%%%%%%%%%%%%%%%%%%%%%%%%%%%%%%%%%%%%%%%%%%%%%%%%%%%%%%%%%%%%%%%%%%%%%
\vspace{0.00cm}
\noindent
{\it Summary}: 
The accurate FFs  of the $f_0 (980)$ were determined 
by a global analysis of the experimental data for the first time.
In particular, the measurements of the Belle collaboration of 2025
on the process $e^+ + e^- \to f_0 (980) +X$ made 
the accurate determination possible. 
From the second moments of the determined FFs, we found
the relation $M_u = M_d \ll M_s \sim  M_g$.
Furthermore, we found that the strange-quark function $D_s(z)$
is distributed in the large-$z$ region in comparison with
the light-quark functions, $D_u(z)$ and $D_d(z)$
and also the gluon function $D_g(z)$.
These relations on the second moments and the functional forms
indicate that $f_0 (980)$ looks like an $s\bar s$ state
from the FFs.
%%%%%
There is also a $s\bar s$-state indication
in high-energy heavy-ion reaction \cite{CMS-f0-2025}.

This fact seems to be contradictory to the common sense 
that $f_0 (980)$ is a tetraquark state or $K\bar K$ molecule
from low-energy studies. The only way to reconcile these
different pictures of the same hadron is that
the internal configuration looks different depending
on the energy we look at the hadron.
Namely, $f_0 (980)$ is  a tetraquark state or $K\bar K$ molecule
at low energies, but it is an $s\bar s$ state at high energies.
This transition picture agrees with the phenomena seen
in the photon production of $\Lambda (1405)$
by using the constituent counting rule of perturbative QCD.

In general, some exotic hadrons may be described by 
the $q\bar q$ and $qqq$ pictures at high energies,
although they could be considered as exotic hadrons 
at low energies.
From this work, we expect that a new field of 
hadron physics could be created for the transition 
from exotic hadrons to the standard $q\bar q$ 
and $qqq$ configurations as the energy increases.

%%%%%%%%%%%%%%%%%%%%%%%%%%%%%%%%%%%%%%%%%%%%%%%%%%%%%%%%%%%%%%%%%%%%%%%%%%%%%%%%
\vspace{0.00cm}
\noindent
{\it Acknowledgements}: 
The author would like to thank suggestions by R. Seidl and M. Takizawa.

\vspace{-0.30cm}
%%%%%%%%%%%%%%%%%%%%%%%%%%%%%%%%%%%%%%%%%%%%%%%%%%%%%%%%%%%%%%%%%%%%%%%%%%%%%%%%

%%%%%%%%%%%%%%%%%%%%%%%%%%%%%%%%%%%%%%%%%%%%%%%%%%%%%%%%%%%%%%%%%%%%%%%%%%%%%%%%

%%%%%%%%%%%%%%%%%%%%%%%%%%%%%%%%%%%%%%%%%%%%%%%%%%%%%%%%%%%%%%%%%%%%%%%%%%%%%%%%

\begin{thebibliography}{99}
\vspace{-0.30cm}
\bibitem{pdg-2024}
   For a recent review on $f_0 (980)$, see
   S. Navas {\it et al}. (Particle Data Group), 
      Phys. Rev. D {\bf 110}, 030001 (2024),
   especially Sec.\,64, 
   Scalar Mesons below 1 GeV, for finding the current
   situation of possible $f_0 (980)$ configurations.
\bibitem{cik-1993}
   F. E. Close, N. Isgur, S. Kumano,
      Nucl. Phys. B {\bf 389}, 513 (1993).
\bibitem{sk-2015}      
   T. Sekihara and S. Kumano,
      Phys. Rev. D {\bf 92}, 034010 (2015).     
\bibitem{hkos-2008} 
   M. Hirai, S. Kumano, M. Oka, and K. Sudoh,
      Phys. Rev. D {\bf 77}, 017504 (2008).
\bibitem{const-conting-2013}
   H. Kawamura, S. Kumano, and T. Sekihara, 
      Phys. Rev. D {\bf 88} 034010, (2013).
\bibitem{const-conting-2016}
   W.-C. Chang, S. Kumano, and T. Sekihara, 
      Phys. Rev. D {\bf 93}, 034006 (2016).
\bibitem{gpd} 
   H. Kawamura and S. Kumano,
      Phys. Rev. D {\bf 89}, 054007 (2014).
\bibitem{FFs-global-1}
   M. Hirai, S. Kumano, T.-H. Nagai, and K. Sudoh,
      Phys. Rev. D 75, 094009 (2007);
   M. Hirai, H. Kawamura, S. Kumano, and K. Saito,  
      Prog. Theor. Exp. Phys. {\bf 2016}, 113B04 (2016). 
\bibitem{FFs-global-2}
   There are many global analyses for light hadrons.
   For example, see the recent works, 
   V. Bertone {\it et al.},
      Eur. Phys. J. C {\bf 78}, 651 (2018);
   R. A. Khalek, V. Bertone, and E. R. Nocera,
      Phys. Rev. D {\bf 104}, 034007 (2021);
   E. Moffat, W. Melnitchouk, T. C. Rogers, and N. Sato
      Phys. Rev. D {\bf 104}, 016015 (2021);
   I. Borsa, D. de Florian, R. Sassot, and M. Stratmann, 
      Phys. Rev. D {\bf 105}, 031502 (2022);
   J. Gao {\it et al.},
      Phys. Rev. D {\bf 110}, 114019 (2024);
      Phys. Rev. Lett. {\bf 132}, 261903 (2024);
      Phys. Rev. Lett. {\bf 135}, 041902 (2025). 
   For a complete list of FFs analyses, see 
   J. Gao {\it et al.},
       Phys. Rev. D {\bf 112}, 054045 (2025).
\bibitem{KEKB-2025}
   R. Seidl {\it et al.}, (Belle Collaboration), 
      Phys. Rev. D {\bf 111}, 052003 (2025).
\bibitem{esw-book} 
    R. K. Ellis, W. J. Stirling, and B. R. Webber,
      {\it QCD and Collider Physics}, Cambridge University Press (1996).
\bibitem{FFs-evol-2012}
   M. Hirai and S. Kumano, 
      Comput. Phys. Commun. {\bf 183}, 1002 (2012).
\bibitem{hrs86}
   S. Abachi {\it et al.} (HRS Collaboration), 
      Phys. Rev. Lett. {\bf 57}, 1990 (1986).    
\bibitem{delphi9599}
   P. Abreu {\it et al.} (DELPHI Collaboration), 
      Z. Phys. C {\bf 65}, 587 (1995);
      Phys. Lett. B {\bf 449}, 364 (1999).
\bibitem{opal98}
   K. Ackerstaff {\it et al.} (OPAL Collaboration), 
      Eur. Phys. J. C {\bf 4}, 29 (1998).   
\bibitem{minuit}
http://www.dnp.fmph.uniba.sk/cernlib/asdoc/minuit/node33.html.
\bibitem{CMS-f0-2025}
   A. Hayrapetyan {\it et al.} (CMS Collaboration), 
      Nature Commu. {\bf 16}, 7990 (2025).
\end{thebibliography}
\end{document}